# Learning and Solving Many-Player Games through a Cluster-Based Representation

**Sevan G. Ficici**
Harvard School of Engineering
and Applied Sciences
sevan@eecs.harvard.edu

**David C. Parkes**
Harvard School of Engineering
and Applied Sciences
parkes@eecs.harvard.edu

**Avi Pfeffer**
Harvard School of Engineering
and Applied Sciences
avi@eecs.harvard.edu

## Abstract

In addressing the challenge of exponential scaling with the number of agents we adopt a *cluster-based representation* to approximately solve asymmetric games of very many players. A cluster groups together agents with a similar "strategic view" of the game. We learn the clustered approximation from data consisting of strategy profiles and payoffs, which may be obtained from observations of play or access to a simulator. Using our clustering we construct a reduced "twins" game in which each cluster is associated with two players of the reduced game. This allows our representation to be *individually-responsive* because we align the interests of every *individual* agent with the strategy of its cluster. Our approach provides agents with higher payoffs and lower regret on average than model-free methods as well as previous cluster-based methods, and requires only few observations for learning to be successful. The "twins" approach is shown to be an important component of providing these low regret approximations.

## 1 Introduction

Consider the problem of solving non-cooperative games of realistic size. As the number of agents increases, the size of the game increases exponentially and it becomes intractable to even represent a game explicitly never mind solve for the equilibrium of the game. Recognizing this, one direction adopted in the literature on computational game theory is to assume that games have some underlying structure and focus on succinctly representable games; e.g., graphical games [6] and action-graph games [2]. But what if there is no exact, structured representation of a game? A second natural direction is to find a suitable approximation of the game, that can be solved and will provide a good model of the strategic characteristics of the actual game, so that agents have low regret from adopting the strategy determined by solving the approximate game. This is the direction we follow in the current paper.[1]

We consider the use of a *cluster-based representation*, in which the same strategy is ultimately prescribed to every agent in a cluster. A cluster groups together agents with a similar "strategic view" of the game. This means that they have similar payoffs and similar effects on other agents. We do not require that the actual game is symmetric and allow for agents with different payoff functions even within a cluster. We do not require explicit knowledge of the full game. Instead, we learn (offline) the cluster-based representation from data consisting of strategy profiles and payoffs. The data may be obtained from observations of play, or we may have access to a simulator with which to generate payoffs for different strategy profiles. We learn both the clustering and also the payoffs to agents in each cluster given each profile of cluster strategies.

A natural next step is to solve a clustered representation of the game to find a Nash equilibrium, and recommend the equilibrium strategies to agents in the original many-player game. However, using a naive clustered representation leads to a situation in which individual agents' interests differ from their clusters', and therefore individual agents will not adhere to the recommended strategies. To address this problem we construct a "twins" game in which every cluster is represented by a pair of players. This representation is *individually-responsive* because in addition to requiring that the cluster strategies form a Nash equilibrium of the reduced game induced by the clusters, we ensure that the interests of every *individual* agent within a cluster are aligned with the cluster strategy. As a result, no individual agent within a cluster would like to deviate from the strategy prescribed to the cluster.

---

[1] Finding approximation algorithms for solving games is an active area (see Daskalakis et al. [3] for a recent survey), but our focus is on effective, heuristic methods rather than on achieving worst-case approximation bounds.

We present a motivating example of a vendor game in Section 2. In Section 3, we introduce our cluster-based model and explain our methods to learn a good clustering from observations of play (or access to a simulator) and to learn a good model of payoffs for a given clustering. In Section 4 we explain how the twins game is defined for a given instantiation of the cluster-based model, and explain why the representation is individually-responsive. We present experimental results in Section 5, where we consider the vendor game (where vendors can be complements or substitutes for each other) and a variant of the *Santa Fe bar problem*. We compare our cluster-based model with other approaches and show that our model provides solutions that give agents low regrets and good payoffs.

**Related Work.** Wellman et al. [11] propose and study an approximate representation that is suitable for symmetric games. To facilitate solving such a game, these authors group agents into multiple clusters, wherein all agents in each cluster are constrained to follow the same strategy. In our approach, we would use a single cluster for a symmetric game and solve it with the twins representation. Our twins approach provides individual responsiveness while allowing for multiple clusters, which is essential for asymmetric games. Furthermore, we do not require that the clustered representation be exact, or that the full game be known, but instead learn the cluster-based representation from observations.

On the other hand, Vorobeychik et al. [10] apply regression learning techniques to model payoffs for continuous games. While these authors consider some forms of strategy aggregation, they do not explicitly seek to combine learning with the use of reduced game-form representations. The novelty of our use of learning then is that we seek to integrate learning directly with our reduced game representation, both for the purpose of learning the structure of an appropriate cluster representation and also for learning the payoffs for the induced "twins" game. Further, our learning is applied to many player games, whereas they restrict attention to games with a small number of players.

The work on graphical games [6] assumes locality of interaction, while we make no such assumptions, and graphical games with many agents remain hard to solve. In comparing with action graph games [2], we note for example that our vendor game could be viewed as an action graph game *if the cluster-based approximation is in fact exact*. But solving the action graph game would still amount to solving a many player game because individual agents would need to take into account other agents in their own cluster. In contrast, our twins game allows us to reduce the number of players while still being individually responsive.

Daskalakis and Papadimitriou [4] consider anonymous games, a special case of action graph games in which the strategic considerations depend only on the *number* of agents that adopt each strategy but not their identity. They develop worst-case approximation results, including a polynomial time approximation scheme for finding an $\epsilon$-approximate Nash equilibrium when the number of strategies is two.

Kearns and Mansour [7] present *summarization games*, a compact representation for games with many players. Their approach has two components: a *summarization function*, which maps the space of $N$-player strategy profiles onto the interval $[0, 1]$, and a set of *payoff functions*, one for each player. Each player's payoff function is a function of that player's strategy choice and the output of the summarization function. They establish conditions under which the Nash equilibria of summarization games can be approximated to within some epsilon in polynomial time, but do not consider the use of summarization games as *approximations* to other games. Indeed, whereas most work on approximation uses an exact representation of the game to obtain an approximate solution, we instead use an approximate representation of the game and obtain an exact solution to that approximation.

Jehiel [5] advances the idea of an *analogy-based expectation* equilibrium in which agents are clustered into "analogy classes" and each agent plays a best-response against the average strategy in each cluster. Whereas our approach solves a reduced twins game with two players per cluster, Jehiel's approach still maintains an explicit representation of every agent in solving for this equilibrium. Rather than computational tractability, Jehiel's focus is on providing prescriptive power for how people behave in strategic settings, along with alternate explanations of some well known paradoxes in extensive form games.

## 2 The Vendor Game

An example of a game that lends itself to our approach is a *vendor game*. Here, we have a large number of vendors, each selling a different product. The products belong to categories. For example, some vendors may sell drinks while others sell sandwiches. Within each category, the products are further differentiated. For example, one drinks vendor may sell beer while another sells lemonade. Each vendor must occupy one of a small number of locations from which to operate her vending services; a vending location may accom-modate more than one vendor. What makes a vendor's decision require strategic thinking is the fact that certain vendors' services are natural *complements* to each other, whereas other services are *substitutes* for each other. Thus, if a sandwich vendor and drink vendor decide to operate in the same location, the two vendors will benefit positively from their complementary relationship. In contrast, if two sandwich vendors decide to operate in the same location, then they will split the customers due to the fact that one vendor is a substitute for the other. It is also possible that two vendors operating in the same location will have no effect on each other's sales; we call this a *neutral* relationship. Vendors operating in different locations have no effect on each other. While these general relationships apply across categories, they differ between vendors in the same category. For example, orange juice may more severely substitute for lemonade than for beer.

In modeling the full vendor game, let $\mathcal{A}$ be the set of vendors (i.e., agents), where $N = |\mathcal{A}|$ is the number of agents. Let $\mathcal{T}$ be the set of product types, where each vendor sells one product type. Let $\mathcal{S}$ be the set of possible vending locations; thus, $\mathcal{S}$ constitutes the set of pure strategies that each vendor has. Let $T$ be a matrix that indicates how product types interact on average. For each possible ordered pair of product types $t_i, t_j \in \mathcal{T}$, $T(i,j)$ specifies the average impact a product of type $t_j$ has on a product of type $t_i$. If $T(i,j) = 0$, then the product types do not interact. If $T(i,j) > 0$, then the two product types are complementary. If $T(i,j) < 0$, then the two product types are substitutes for each other.

The matrix $T$ describes how product types interact on average. Let matrix $A$ describe how two specific vendors interact, in particular; $A(x,y)$ gives the impact that vendor $a_y$ has on vendor $a_x$. Let vendors $a_x, a_y \in \mathcal{A}$ sell products of types $t_i, t_j \in \mathcal{T}$, respectively; the value of $A(x,y)$ is obtained by sampling once from the normal distribution $N(T(i,j), \sigma^2)$, where $\sigma^2$ is a game parameter. In addition to the agent interactions, each agent $a_x$ has an associated bias term that specifies a base-line success level for the agent independent of the impacts of other agents. Let $b_x$ be the bias for agent $a_x$.

A pure strategy profile in the game is associated with the selection of a location by each agent. Let $s_x$ represent the pure strategy choice of agent $a_x$. The payoff to player $a_x \in \mathcal{A}$ playing strategy $s_x \in \mathcal{S}$ is:

$$\pi_{a_x}(s_x, s_{-x}) = b_x + \sum_{a_y \in \mathcal{A}} \begin{cases} A(x,y) & \text{if } s_x = s_y \\ 0 & \text{if } s_x \neq s_y \end{cases} \quad (1)$$

Because the interaction between each pair of vendors is different, this is a many player game with no obviously exploitable structure in its exact representation. One solution method would be to construct a normal-form representation of the game and apply a standard solution algorithm. However, the size of the normal form representation is exponential in the number of vendors, so this method quickly becomes infeasible. Our approach is based on the observation that in this game, the agents naturally fall into clusters. Vendors in the same category have a similar strategic view of the game. They tend to be affected in similar ways by other vendors, and they also affect other vendors in similar ways. We next expand on this intuition.

## 3 The Cluster-Based Model

Our approach to compactly representing and tractably solving asymmetric $N$-player games is to use a (generally) lossy compression. We group a large number of agents into a much smaller number of clusters. The fundamental assumptions made by our approximation are that 1) agents clustered into the same cluster receive similar payoffs when they take the same action;[2] 2) agents clustered into the same cluster have similar influences on other agents in the same and different clusters; and 3) within each cluster, the actions of the individual agents are combined linearly. Thus, the combinatorial effects of strategic interaction are captured at the level of the cluster, which is how we realize our computational savings. In this section we introduce the cluster-based representation and explain how the model (both the cluster structure and the payoffs) is learned from data. Before continuing, let us remark that while we allow our game to be asymmetric (in payoffs) we do presently require each agent to share the same set of pure strategies $\mathcal{S}$.

### 3.1 Defining the Model

Let $K$ be the number of clusters that we wish to generate; this parameter allows us to express a trade-off between fidelity to the original game and computational efficiency. Let $\mathcal{C}$ be a clustering of the $N$ agents into $K$ clusters, with $C_i \in \mathcal{C}$ being the $i$th cluster of agents. A cluster will be termed a "player" in the reduced game induced by a clustering. We reserve the term "agent" for an agent of the original $N$-player game.

For each combination of a cluster and a pure strategy (i.e., each element in the cross product $\mathcal{C} \times \mathcal{S}$), we construct a linear equation with $|\mathcal{S}|^K + 1$ terms. Each one of these linear equations $\hat{\pi}_{C_i}(s_x)$ is a regressor that estimates the payoff an agent in cluster $C_i$ would receive when it plays pure strategy $s_x$, given the clustering $\mathcal{C}$ and probabilistic information about the strategy profile adopted by each cluster. Note that strategy $s_x$ may be different from the strategy adopted by the cluster with which agent $a_x$ is associated. The fact that there is one equation per cluster per strategy captures the first assumption that all agents in a cluster receive the same payoff for the same action.

---
[2] If necessary, agents' payoffs can be normalized, according to the data, to bring payoffs into the same scale.

One of the terms in $\hat{\pi}_{C_i}(s_x)$ is simply a constant $\beta_{C_i(s_x)}$ to capture any "offset" effects. In defining the remaining terms, let $\vec{s} = (s_1, \ldots, s_K)$ denote a strategy adopted by each cluster. Given this, then the remaining terms are of the form $\beta^{\vec{s}}_{C_i(s_x)} \Pr(\vec{s}|\mathcal{C})$, which is the product of a constant $\beta^{\vec{s}}_{C_i(s_x)}$ and the estimated probability $\Pr(\vec{s}|\mathcal{C})$ with which cluster strategy profile $\vec{s}$ is adopted by the underlying agents, given empirically observed data about the game (for all $N$ agents) and the clustering $\mathcal{C}$. More precisely, assuming the strategies of different agents are uncorrelated, we estimate the joint probability as

$$\Pr(\vec{s}|\mathcal{C}) = \Pi_{i=1}^{K} \Pr(s_i|C_i), \qquad (2)$$

where $\Pr(s_i|C_i)$ is the estimated probability that an agent in cluster $C_i$ plays pure strategy $s_i$. This probability is estimated from the proportion of agents in the cluster that play strategy $s_i$. Thus, the payoff to an agent in a cluster depends only on the proportion of agents playing each strategy in each cluster, and not on the actions of individual agents. This captures our second assumption above. The fact that the equations are linear captures our third assumption.

**Example.** To concretely illustrate our regressor equations, let us assume a game where all agents have two pure strategies, $L$ and $R$, and we wish to cluster agents into two clusters, $A$ and $B$. We thus generate four regressors. For example, the estimated payoff obtained by an agent in cluster $A$ playing strategy $L$ is denoted $\hat{\pi}_A(L)$. Following this notation, the payoffs for agents in clusters $A$ and $B$, using strategies $L$ and $R$, are:

$$\hat{\pi}_A(L) = \beta^{L,L}_{A(L)} \cdot \Pr(L|A) \cdot \Pr(L|B) +$$
$$\beta^{L,R}_{A(L)} \cdot \Pr(L|A) \cdot \Pr(R|B) + \beta^{R,L}_{A(L)} \cdot \Pr(R|A) \cdot \Pr(L|B) +$$
$$\beta^{R,R}_{A(L)} \cdot \Pr(R|A) \cdot \Pr(R|B) + \beta_{A(L)} \qquad (3)$$

$$\hat{\pi}_A(R) = \beta^{L,L}_{A(R)} \cdot \Pr(L|A) \cdot \Pr(L|B) +$$
$$\beta^{L,R}_{A(R)} \cdot \Pr(L|A) \cdot \Pr(R|B) + \beta^{R,L}_{A(R)} \cdot \Pr(R|A) \cdot \Pr(L|B) +$$
$$\beta^{R,R}_{A(R)} \cdot \Pr(R|A) \cdot \Pr(R|B) + \beta_{A(R)} \qquad (4)$$

$$\hat{\pi}_B(L) = \beta^{L,L}_{B(L)} \cdot \Pr(L|A) \cdot \Pr(L|B) +$$
$$\beta^{L,R}_{B(L)} \cdot \Pr(L|A) \cdot \Pr(R|B) + \beta^{R,L}_{B(L)} \cdot \Pr(R|A) \cdot \Pr(L|B) +$$
$$\beta^{R,R}_{B(L)} \cdot \Pr(R|A) \cdot \Pr(R|B) + \beta_{B(L)} \qquad (5)$$

$$\hat{\pi}_B(R) = \beta^{L,L}_{B(R)} \cdot \Pr(L|A) \cdot \Pr(L|B) +$$
$$\beta^{L,R}_{B(R)} \cdot \Pr(L|A) \cdot \Pr(R|B) + \beta^{R,L}_{B(R)} \cdot \Pr(R|A) \cdot \Pr(L|B) +$$
$$\beta^{R,R}_{B(R)} \cdot \Pr(R|A) \cdot \Pr(R|B) + \beta_{B(R)} \qquad (6)$$

Since we have two agent clusters and two pure strategies, we have four regressors, each with five terms. For example, in Equation (3) we are estimating the payoff received by an agent in cluster $A$ when it plays strategy $L$. Note that the third term, $\beta^{R,L}_{A(L)} \cdot \Pr(R|A) \cdot \Pr(L|B)$, represents the contribution to our agent's payoff that occurs when agents in cluster $B$ play $L$ in combination with agents in cluster $A$ playing $R$. Thus, even though the agent for whom we are calculating the payoff is itself playing $L$, we are accounting for the effect that any other agents within cluster $A$ who are playing $R$ (in combination with cluster $B$ agents playing $L$) may have upon our agent's payoff.

### 3.2 Model Learning

A key aspect to our work is that we do not require *a priori* knowledge of the cluster-based model. Instead, we may learn the model from observation of agent behavior and earned payoffs. We have two distinct settings in mind:

• We have access to a data set about agent actions and agent payoffs in the underlying game.

• We have access to a simulator that we can use to generate payoffs for different strategy profiles in the underlying game.

In both cases we are learning the model offline and adopting the viewpoint of an analyst, not an agent situated within the strategic environment. Notice that in the second approach we need not have a complete representation of the game, which would in general be too large to represent. Rather, all that is required is a way to generate payoffs for different strategy profiles.

From our observations we learn our cluster-based model of the underlying $N$-player game. Each observation consists of an $N$-tuple pure strategy profile and the corresponding payoffs received by each agent. We require that every action in every cluster be seen at least once in order to learn.

Given a clustering of the $N$ agents into $K$ clusters, we can learn the $\beta$ parameters for our regressors with linear regression. Each observation of an $N$-agent interaction typically provides several data instances for the linear regression. For instance, consider our example game above. Let cluster $A$ contain $N_A$ agents, each playing a uniform mixed strategy over pure strategies $L$ and $R$. If we have $M$ observations, then we expect to obtain $N_A M/2$ data instances each for regressor Equations (3) and (4). In this way, each additional agent that we might place into our full game will linearly increase the number of instances for the linear regression. Thus, we expect the number of observations of the full game that we need to be inversely proportional to the number of agents, at least for the purpose of estimating parameters.

**Agent Clustering.** We may compute a linear regression given any clustering of $N$ agents into $K$ clusters. Clearly, we desire the clustering $\mathcal{C}$ that provides us with the best regressors, so that we can most accurately estimate agent payoffs. We use the sum of squared errors over all agents over all regressor equations to quantify the quality of our estimations, where the error of a regressor for an agent is the difference between the estimated payoff for the agent, given the $N$-tuple strategy profile, and the agent's actual payoff.

Since we assume a finite number of agents, there exist a finite number of possible clusterings into $K$ clusters; an optimal clustering must therefore exist. Nevertheless, the large number of possible clusterings precludes exhaustive search. Therefore, we use $k$-means clustering to obtain the clustering $\mathcal{C}$ that we use to run our linear regressions.

A key intuition in our work is that agents clustered into the same cluster receive relatively similar payoffs when they take the same action, given a shared context of what the other agents in the system do. We use this intuition to construct the features with which $k$-means operates. Specifically, we place each agent in an $S$-dimensional space, where $S$ is the number of pure strategies available to agents. Each dimension corresponds to a pure strategy, and the location of an agent in a particular dimension is the average payoff the agent earned over our observations when it used the corresponding pure strategy. Note that, here, we are not conditioning further on what other agents do; this is to keep both dimensionality and computational complexity low. Because $k$-means does not always converge onto the same clustering each time it is run, we run it several times and choose the result that gives us the lowest sum of squared errors in our regressors.

## 4 The Twins Game

Given that we have $K$ clusters, an obvious reduction would be to construct a $K$-player normal form game. We would then solve this smaller $K$-player game and then assign each agent within cluster $i$ the strategy (pure or mixed) used in Nash equilibrium by the $i$-th player in the $K$-player game. This would follow the approach of Wellman et al. [11], albeit slightly extended to an asymmetric setting. But one problem with this approach is that, while the $i$-th player in the $K$-player game has no incentive to unilaterally deviate from equilibrium, cluster $i$ cannot be treated as a true "player" because it is not a monolithic decision maker. Each cluster of agents consists of independent decision makers, whose individual incentives might not be aligned with the cluster-level incentives of the $K$-player game. Thus, an individual in some cluster may, in fact, wish to deviate from the prescribed $K$-player strategy profile.

Rather than build a $K$-player game, we build a $2K$-player game, where each cluster is represented twice, hence what we call a "twins game." Each cluster $C$ is associated with two players, $\mathbb{C}$ and $\mathbb{C}'$. (Recall that a player captures a strategic entity in the clustered game and is distinct from an agent in the underlying game.) These players have multiple interpretations, depending on from whose point of view the players are being considered. When considering the payoff of player $\mathbb{C}$, $\mathbb{C}$ is interpreted as representing a single agent in cluster $C$, while $\mathbb{C}'$ represents the rest of the agents in the cluster in aggregate. The view from $\mathbb{C}'$ is symmetric: the payoff of player $\mathbb{C}'$ represents that of a single agent in the cluster, while $\mathbb{C}$ represents the rest of the cluster $C$ in aggregate. From the point of view of a player $\mathbb{D}$ associated with a different cluster $D \neq C$, $\mathbb{C}$ and $\mathbb{C}'$ together represent cluster $C$. When $\mathbb{C}$ and $\mathbb{C}'$ adopt different strategies $s_\mathbb{C}$ and $s_{\mathbb{C}'}$, then for symmetry we consider that half of the agents in the cluster play $s_\mathbb{C}$ and half play $s_{\mathbb{C}'}$.

Under these interpretations, we can obtain the payoffs for each of the $2K$ players in the twins game for a given strategy profile. Consider the player $\mathbb{C}$ associated with cluster $C$, where $\mathbb{C}$ plays strategy $s_\mathbb{C}$ and $\mathbb{C}'$ plays strategy $s_{\mathbb{C}'}$. We instantiate the probabilities for the other players in the linear regression model, and thus define the payoff to player $\mathbb{C}$, as follows,

$$\Pr(s_i|C) = \begin{cases} 1 & \text{if } s_i = s_{\mathbb{C}'} \\ 0 & \text{otherwise} \end{cases}$$

For a cluster $D \neq C$,

$$\Pr(s_i|D) = \begin{cases} 1 & \text{if } s_i = s_\mathbb{D} = s_{\mathbb{D}'} \\ 1/2 & \text{if one of } s_\mathbb{D} \text{ or } s_{\mathbb{D}'} = s_i \\ 0 & \text{otherwise} \end{cases}$$

where $\mathbb{D}$ and $\mathbb{D}'$ are the players corresponding to cluster $D$. The payoff to $\mathbb{C}$ is then $\hat{\pi}_C(s_\mathbb{C})$ as specified by the regressor equations. The payoff to $\mathbb{C}'$ is symmetric.

By representing each cluster twice, we can seek Nash equilibria in which the incentives of individuals within a cluster are aligned with the cluster itself. We care to locate Nash equilibria where each player and its twin use the same strategy. We will call such equilibria *twin symmetric* Nash equilibria (TSNE). A TSNE is guaranteed to exist, because there is always a twin symmetric best response to a twin symmetric strategy profile, so we can use the same argument as in Nash's proof of the existence of Nash equilibrium [9].

Let $s_\mathbb{C}$ denote this strategy (pure or mixed), used by both players $\mathbb{C}$ and $\mathbb{C}'$ in a TSNE (and in turn used by each underlying agent within the cluster.) From the perspective of player $\mathbb{C}$, strategy $s_\mathbb{C}$ is a best response to the strategies of the other players, and particularly to $\mathbb{C}'$ who also plays $s_\mathbb{C}$. In the construction of the twins game, the payoff to player $\mathbb{C}$ equals the payoff to an individual agent in cluster $C$ playing $s_\mathbb{C}$ when the rest of cluster $C$ plays $s_{\mathbb{C}'}$. But in a TSNE, $s_\mathbb{C} = s_{\mathbb{C}'}$. Thus, if $s_\mathbb{C}$ is the strategy recommended to $\mathbb{C}$ in a TSNE, an individual agent in cluster $C$ will be playing a best response to its cluster as a whole playing $s_\mathbb{C}$. In this way, we say that our representation is *individually-responsive*. No individual agent will have incentive to deviate from the strategy recommended to its cluster.

### 4.1 Example

Let us continue our example from Section 3.1. We have two clusters of agents, $A$ and $B$, and so will build a four player game. Let us label the players $\mathbb{A}$, $\mathbb{A}'$, $\mathbb{B}$, and $\mathbb{B}'$, where $\mathbb{A}$ and $\mathbb{A}'$ correspond to the twin players for cluster $A$, and similarly for the players for cluster $B$. Table 1 shows how we convert our regressors to a twins game. Each row of the table corresponds to a single four-player pure-strategy profile. For brevity, we show only 1/4 of the profiles, showing only those where $\mathbb{A}$ plays $L$ and $\mathbb{A}'$ plays $R$. The leftmost column specifies a four-player pure-strategy profile; for example, the third row specifies a profile where players $\mathbb{A}$, $\mathbb{A}'$, $\mathbb{B}$, and $\mathbb{B}'$ play pure strategies $L$, $R$, $R$, and $L$, respectively. The remaining columns indicate the payoffs received by each player for each pure-strategy profile.

We compute player payoffs as described in the previous section. The pure strategy that a player uses in a profile determines which regressor equation we will use. For example, in the profile $LRLL$, player $\mathbb{A}$ plays $L$, and so we use Equation (3); player $\mathbb{A}'$ plays $R$, and so for this player we use Equation (4). We use Equation (5) for players $\mathbb{B}$ and $\mathbb{B}'$, since both play $L$.

The payoff to $\mathbb{A}$ for the profile $LRLL$ is $\beta_{A(L)}^{RL} + \beta_{A(L)}$. This is the payoff that a cluster $A$ agent would receive if it played $L$ in the situation where all cluster $A$ agents actually play $R$ (i.e., $\Pr(R|A) = 1.0$) and all cluster $B$ agents play $L$. Similarly, the payoff to $\mathbb{A}'$ is $\beta_{A(R)}^{LL} + \beta_{A(R)}$, which corresponds to the payoff a cluster $A$ agent would receive for playing $R$ in the situation where all cluster $A$ agents actually play $L$ and all cluster $B$ agents play $L$. More interesting is the payoff for $\mathbb{B}$, which is $\frac{\beta_{B(L)}^{LL} + \beta_{B(L)}^{RL}}{2} + \beta_{B(L)}$. This is the payoff obtained by a cluster $B$ agent if it would play $L$ in the situation where all cluster $B$ agents actually play $L$ and where half the cluster $A$ agents play $L$ and half play $R$ (i.e., $\Pr(L|A) = \Pr(R|A) = 0.5$). Thus, when a player and its twin (e.g., $\mathbb{A}$ and $\mathbb{A}'$) play different pure strategies in a profile, we interpret this as a uniform distribution over the two strategies when computing the payoff for another player (e.g., $\mathbb{B}$ or $\mathbb{B}'$) in the twins game.

## 5 Experimental Results

In generating our observation set for the purpose of experimentation, we build $N$ agents to play the game for some number of interactions. We provide agents with a uniform distribution over their pure strategies in order to generate an observation set with good support in the space of possible joint strategy profiles in our twins game. From these observations, and with a given number $K$ of clusters, we learn our cluster-based approximation; this entails clustering and linear regression. We then construct the $2K$-player twins game. For comparison, we also construct a $K$-player game, with one player per cluster and thus without individual responsiveness. Using Gambit [8], we find all Nash equilibria of these reduced normal-form games. For each TSNE of the $2K$-player game and each NE of the $K$-player game, we assign the equilibrium strategies to the agents and have them interact for 100 iterations. We then calculate mean payoffs and external regret values for each agent in this new data. Note that we do not need to solve the full $N$-player game.

In addition to comparing results from the $2K$- and $K$-player games, we compare the performance of our approach with that of two model-free learning approaches. These methods are a form of reinforcement learning whereby agents play strategies that have yielded the highest mean rewards. Our first model-free approach concerns *agent-level learning* (ALL). We examine the initial observation set to determine, for each agent, which pure strategy provided the agent with the highest mean payoff. Each agent then adopts the pure strategy that provided it the highest mean payoff for use in future interactions. We generate new interaction data with agents playing these pure strategies, and then determine mean agent payoff and regret values for the new data. Our second model-free approach provides *cluster-level learning* (CLL). Here we examine the initial observation set to determine, for each

Table 1: Conversion of regression equations to normal-form "twins game" (partial specification).

| Profile | $\mathbb{A}$ | $\mathbb{A}'$ | $\mathbb{B}$ | $\mathbb{B}'$ |
|---|---|---|---|---|
| $LRLL$ | $\beta_{A(L)}^{RL} + \beta_{A(L)}$ | $\beta_{A(R)}^{LL} + \beta_{A(R)}$ | $\frac{\beta_{B(L)}^{LL}+\beta_{B(L)}^{RL}}{2} + \beta_{B(L)}$ | $\frac{\beta_{B(L)}^{LL}+\beta_{B(L)}^{RL}}{2} + \beta_{B(L)}$ |
| $LRLR$ | $\frac{\beta_{A(L)}^{RL}+\beta_{A(L)}^{RR}}{2} + \beta_{A(L)}$ | $\frac{\beta_{A(R)}^{LL}+\beta_{A(R)}^{LR}}{2} + \beta_{A(R)}$ | $\frac{\beta_{B(L)}^{LR}+\beta_{B(L)}^{RR}}{2} + \beta_{B(L)}$ | $\frac{\beta_{B(R)}^{LL}+\beta_{B(R)}^{RL}}{2} + \beta_{B(R)}$ |
| $LRRL$ | $\frac{\beta_{A(L)}^{RL}+\beta_{A(L)}^{RR}}{2} + \beta_{A(L)}$ | $\frac{\beta_{A(R)}^{LL}+\beta_{A(R)}^{LR}}{2} + \beta_{A(R)}$ | $\frac{\beta_{B(R)}^{LL}+\beta_{B(R)}^{RL}}{2} + \beta_{B(R)}$ | $\frac{\beta_{B(L)}^{LR}+\beta_{B(L)}^{RR}}{2} + \beta_{B(L)}$ |
| $LRRR$ | $\beta_{A(L)}^{RR} + \beta_{A(L)}$ | $\beta_{A(R)}^{LR} + \beta_{A(R)}$ | $\frac{\beta_{B(R)}^{LR}+\beta_{B(R)}^{RR}}{2} + \beta_{B(R)}$ | $\frac{\beta_{B(R)}^{LR}+\beta_{B(R)}^{RR}}{2} + \beta_{B(R)}$ |

*cluster*, which pure strategy provided the agents in the cluster with the highest mean payoff. Each agent in the cluster then adopts this pure strategy. We then generate new interaction data.

### 5.1 Vendor Game

We initialize the type-interaction matrix $T$ as follows. Each product type is treated as a substitute for itself, thus the diagonal of $T$ is negative. Off the diagonal, we randomly select between interaction types ($\Pr(\text{neutral}) = 0.1, \Pr(\text{substitute}) = \Pr(\text{complement}) = 0.45$), but require $T$ to contain at least one complementary interaction. Substitution means are drawn uniformly between $[-3.0, 0.0]$, complements from $[0.0, 3.0]$, and neutral interactions have a mean of zero.

In one experiment, we use 100 agents, two agent types, two locations ($L$ and $R$), $\sigma^2 = 1.5$, and 15 observations per trial over ten trials. We cluster agents into two groups ($A$ and $B$). Table 2 summarizes our experiment, showing mean agent payoffs and regrets. The model-free learning methods clearly perform the worst; the strategies (always pure) they learn give agents the worst mean payoffs and highest mean regrets. The differences between the $K$- and $2K$-player games are most pronounced with respect to regret levels. When we examine separately the performance of pure-strategy Nash equilibria (PSNE) and mixed-strategy NE (MSNE) of the $K$- and $2K$-player games, we find that agents' regret levels when they use MSNE derived from $K$-player games are over 2.3 times as high as when they use mixed TSNE from $2K$-player games ($p < 0.004$). The $K$-player games often produce PSNE, and the regret levels from these PSNE are an order of magnitude higher than the TSNE of the $2K$-player game. This shows that our twins-game approach aligns the incentives of individual agents.

In another experiment, we look at different numbers of agents ($N = 10$, 100, or 200) combined with differ-ent numbers of observations (3, 5, 10, or 15); other parameters are as above. Table 3 gives average regrets for some of our settings. Looking over all of the data, two trends appear to emerge. First, the disparities between the approaches appears to grow with the number of agents; thus, it becomes more important to have a good strategy in our game with more agents. Second, the level of regret for a given number of agents appears to consistently diminish with our approach as the number of observations increases. Nevertheless, our method consistently finds solutions that give the lowest average regret over all settings. We also ran experiments where we varied the level of noise $\sigma^2$ in agent interactions. Though the $R^2$ values of our regressions suffer, the performance of our approach is surprisingly robust. The general conclusion is that the model-free methods perform poorly, and that the twins game approach produces lower agent regret values than the game-form with one player per cluster.

Table 2: Results from vendor game. 100 agents, 2 types, $\sigma^2 = 1.5$, $K = 2$, 15 observations per trial. When analysis gives multiple NE in a $K$- or $2K$-player game, we select the NE giving the worst result. First row: mean payoffs over all agents over all trials. Significant differences are CLL vs. ALL ($p < 0.065$ paired sign-rank test), CLL and ALL vs. $K$- and $2K$-player games ($p < 0.014$). Second row: mean external regret over all agents over all trials. Significant differences are CLL vs. ALL ($p < 0.11$), CLL and ALL vs. $K$ and $2K$ ($p < 0.01$), and $K$ vs. $2K$ ($p < 0.04$).

|  | CLL | ALL | $K$-Player | $2K$-Player |
|---|---|---|---|---|
| Payoff | -53.98 | -40.20 | -17.69 | **-15.50** |
| Regret | 87.75 | 64.44 | 17.56 | **3.06** |

### 5.2 Santa Fe Bar

The second game on which we test our approach is a slight variation of the *Santa Fe bar problem* (also known as the *El Farol* bar problem) [1]. In this game, each of the $N$ agents must independently de-cide whether or not to visit the *El Farol* bar. Unfortunately, the bar's capacity $c \in (0, 1)$ allows only $\lfloor cN \rfloor$ agents to fit comfortably. The set of pure strategies for each agent consists of two items: either the agent visits the bar, or stays at home. There are three possible outcomes for an agent: 1) the agent decides to visit the bar, but the bar is full (denoted $v^\bullet$), 2) the agent decides to visit the bar, and the bar has room (denoted $v^\circ$), 3) the agent decides to stay at home (denoted $h$). Each agent prefers these outcomes as follows: $v^\circ \succ v^\bullet$, $v^\circ \succ h$, and $h \succ v^\bullet$. Once all $N$ agents make their choice to visit the bar or stay at home, we obtain an $N$-player pure-strategy profile. The utility for agent $a_x$ for outcome $o$ is denoted by $U_x(o)$.

Table 3: Average regrets from vendor game for different numbers of agents and observations. First column indicates numbers of agents and observations. Note that neither agent payoffs nor regret values are normalized by $N$.

|         | CLL      | ALL      | $K$-Pl.  | $2K$-Pl. |
|---------|----------|----------|----------|----------|
| 10, 3   | 7.6767   | 7.5127   | 4.2450   | 2.9837   |
| 10, 5   | 8.3199   | 6.1706   | 3.3321   | 0.5177   |
| 10, 10  | 7.8606   | 5.9436   | 0.7290   | 0.2226   |
| 100, 3  | 94.2589  | 32.4916  | 15.6744  | 4.3206   |
| 100, 5  | 85.0618  | 69.1498  | 10.5996  | 1.3062   |
| 200, 3  | 181.1111 | 122.8304 | 33.5588  | 5.4801   |
| 200, 15 | 151.1831 | 104.2937 | 2.0577   | 0.9401   |

Table 4: Average regrets from Santa Fe bar game with different bar capacities. For capacity $c = 0.4$, the difference in regret between our method and W$\downarrow_5$ is not statistically significant. For capacity $c = 0.5$, the differences in regret between our method, W$\downarrow_2$ and W$\downarrow_5^*$ are not statistically significant. All other differences between our approach and the others are statistically significant ($p < 0.006$ paired sign-rank test).

| $c$ | $2K$   | W$\downarrow_2$ | W$\downarrow_5$ | W$\downarrow_2^*$ | W$\downarrow_5^*$ |
|-----|--------|-----------------|-----------------|-------------------|-------------------|
| 0.4 | 0.4820 | 4.0             | 0.3508          | 4.0               | 0                 |
| 0.5 | 0.7264 | 0.7750          | 1.6908          | 0                 | 0.6690            |
| 0.6 | 0.8368 | 1.8112          | 0.2068          | 1.8080            | 0                 |

Each agent shares the same utility function, giving us a symmetric game. Since agents all share the same preferences and affect others in the same way, we cluster all agents into a single group; one cluster yields a two-player twins game to solve. Thus, this game exactly satisfies the first two assumptions of our cluster-based representation. Note, however, that it violates the third assumption: the payoffs are not linear in the number of agents playing each strategy. Unlike in the vendor game, an agent's payoff here is only one of three possible values, conditioned on whether the agent visits the bar or stays at home, and whether the total attendance exceeds the bar's capacity or not. The degree to which capacity is exceeded, or the amount of available space, does not affect an agent's payoff.

Because the Santa Fe bar game is symmetric, we can use the method of Wellman, et al., [11] (WEL) to approximate a solution. Their approach to compressing a symmetric game is to "coarse code" the profile and outcome space. They divide the agents into some number of equally sized groups. Given the constraint that all agents within a group must play the same strategy, they perform equilibrium computations using the subset of realizable strategy profiles and outcomes of the *original N-player game*. Since we compress the Santa Fe game down to a two-player twins game, the most direct comparison with WEL is to use their approach to divide the $N$ agents into two groups of $N/2$ agents each, which also yields a two-player game. This approach is denoted by W$\downarrow_2$. However, the approach of WEL allows for more refined approximations that use more clusters. As a point of comparison, we also consider their approach with five clusters, denoted by W$\downarrow_5$, which involves solving a five-player game.

Table 4 summarizes regret values obtained in experiments that examine different bar capacities $c \in \{0.4, 0.5, 0.6\}$. For all three capacity values, we set agent payoffs to be $U(v^\circ) = 4.0$, $U(v^\bullet) = -6.0$, and $U(h) = 0.0$; we use ten agents. We run ten trials for each value of bar capacity. We use 45 observations (agent interactions) per trial for learning in our approach. For complete comparison, we show the regret of the symmetric MSNE found by WEL (shown as W$\downarrow_2$ and W$\downarrow_5$), and also the regret of the best NE found by WEL, whether symmetric or asymmetric (shown as W$\downarrow_2^*$ and W$\downarrow_5^*$). Results for ALL and CLL are not shown in this table; they are significantly worse than the other approaches.

When the bar's capacity is exactly a multiple of the fraction of agents within a cluster, WEL discovers asymmetric PSNE that are actually NE of the full game. Except for this case, we see that our method is better than W$\downarrow_2$ when the capacity is $c = 0.4$ or $c = 0.6$, and is statistically equivalent to it when $c = 0.5$. Even when we compare to WEL allowing it to use five clusters, our method performs reasonably well. It is statistically equivalent to W$\downarrow_5$ when $c = 0.4$. When $c = 0.5$, it is statistically equivalent to the asymmetric equilibrium found by WEL (W$\downarrow_5^*$), and significantly better than the symmetric equilibrium (W$\downarrow_5$). However when $c = 0.6$, W$\downarrow_5$ is significantly better. Note that our results are consistently good despite the fact that our approach involves the added step of learning (which WEL does not).

Figure 1 shows the actual symmetric MSNE for our Santa Fe game for bar capacities $c = 0.4, 0.5, 0.6$. The figure also shows the approximations to these mixed Nash equilibria obtained with our method as well as with $W\downarrow_2$ and $W\downarrow_5$. We see that our approach is more consistently close to the actual symmetric MSNE of the Santa Fe game. Thus, our approach appears better able to detect changes in bar capacity, while at the same time requiring that only a two-player twins game be solved ($W\downarrow_2$ and $W\downarrow_5$ require two- and five-player games to be solved). Indeed, it can be show analytically that the finest distinction $W\downarrow_2$ can make is whether bar capacity is less than $N/2$ or not.

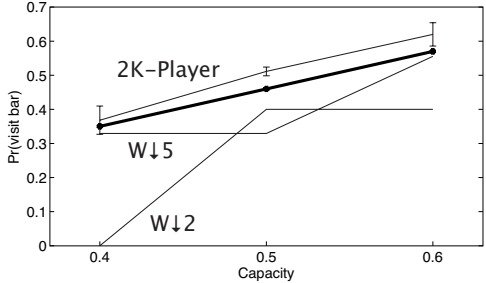

Figure 1: Symmetric mixed strategies obtained for Santa Fe bar game for bar capacities $c = 0.4, 0.5, 0.6$. X-axis indicates bar capacity, Y-axis indicates the probability with which each agent visits the bar in the mixed strategy. Data labeled "2K-Player" indicate the average mixed Nash equilibria (and standard deviation) of our 2K-player twins game over all 10 trials. Data labeled "$W\downarrow_2$" and "$W\downarrow_5$" indicate the mixed Nash equilibria obtained with WEL using two and five clusters, respectively. Bold line indicates the actual mixed Nash equilibria of the Santa Fe game.

## 6 Conclusions

We present a method for approximating the structure of asymmetric $N$-player games for large $N$. This approximation uses a clustering approach to compress the original $N$-player game into a vastly smaller and more tractably solved $2K$-player game, where $K$ is the number of groups into which we cluster the $N$ agents. Each of the $K$ groups of agents is associated with two players in the $2K$-player game; we call these player pairs "twins." Nash equilibria in which twins use the same strategy are "twin-symmetric" NE. We treat these NE as $K$-player strategy profiles that we assign to our agent groups; all agents within a group play the same strategy. The $K$-player NE derived from twin-symmetric NE have the property that the incentives of all agents within a cluster are aligned with the strategy assigned to that cluster. This prevents unilateral deviation by individual agents from the $K$-player NE. Importantly, our method does not assume knowledge of the full $N$-player payoff function; instead, we can learn an approximation to the payoff function through a small number of observations of agent interaction.

We test our method on two different types of game: a vendor game and the Santa Fe bar problem. We show that when agents play solutions obtained from our method, they achieve higher average payoffs and lower external regret compared with two model-free learning approaches we examine. We also compare to compression results obtained with a modified version of our approach that omits the $2K$-player game. We show that this variation is significantly less effective in providing low regret. For the Santa Fe bar problem, data show that our method provides low regret values more consistently than the compression method by Wellman, et al. [11] when using the same number of agent clusters. We also note that our approach is orthogonal to that of Wellman, et al. [11]; for example, we can divide the clusters of the twins game into subclusters to allow for asymmetric equilibria, even in symmetric games and even if learned with just one cluster.

## Acknowledgments

The authors thank the anonymous reviewers for their helpful comments. The research reported in this paper was supported in part by AFOSR grant FA9550-05-1-0321 and NSF grant DMS 0631636.